\documentclass[10pt,conference]{IEEEtran}
\IEEEoverridecommandlockouts
\usepackage{cite}
\usepackage{array}

\usepackage{amsmath,amssymb,amsfonts}
\usepackage{algorithmic}
\usepackage{graphicx}
\usepackage{textcomp}
\usepackage{xcolor}
\usepackage{xspace}

\usepackage[printwatermark]{xwatermark}

\newcommand{\ie}[0]{{\em i.e.},\xspace}
\newcommand{\vs}[0]{{\em vs.}\xspace}
\newcommand{\eg}[0]{{\em e.g.},\xspace}

\def\BibTeX{{\rm B\kern-.05em{\sc i\kern-.025em b}\kern-.08em
 T\kern-.1667em\lower.7ex\hbox{E}\kern-.125emX}}
\begin{document}

\title{The Trip to The Enterprise Gourmet Data Product Marketplace through a Self-service Data Platform}

\author{\IEEEauthorblockN{Micha{\l} Zasadzi{\'n}ski, Michael Theodoulou, Markus Thurner and
Kshitij Ranganath}
\IEEEauthorblockA{Cimpress Technology - USA, Spain, Switzerland \{michal.zasadzinski, mthurner, kshitij.ranganath\}@cimpress.com}  michael.theodoulou@gmail.com}

\maketitle
\thispagestyle{plain}
\pagestyle{plain}

\begin{abstract}
Data Analytics provides core business reporting needs in many software companies, acts as a source of truth for key information, and enables building advanced solutions, \eg predictive models, machine learning, real-time recommendations, to grow the business. Typically, companies ingest data to data lakes, have elaborate ETL/ELT processes, and consume the data. Organizations have challenges in becoming data-driven at scale, with centralized teams and modules acting as bottlenecks, responsible for many aspects such as data governance, ingestion, or engineering. Placing the tools in the hands of developers and empowering engineers across all business domains truly unlocks the ability for a company to become data-driven, at scale, both in culture and practice.

A self-service, multi-tenant, API-first, and scalable \textit{data platform} is the foundational requirement in creating an enterprise data marketplace, which enables the creation, publishing, and exchange of data products. Such a marketplace enables the exploration and discovery of data products, further providing high-level data governance and oversight on marketplace contents.

In this paper, we describe our way to the \textit{gourmet data product marketplace}. We cover the design principles, the implementation details, technology choices, and the journey to build an enterprise data platform that meets the above characteristics. The platform consists of ingestion, streaming, storage, transformation, schema generation, fail-safe, data sharing, access management, PII data automatic identification, self-service storage optimization recommendations, and CI/CD integration. We then show how the platform enables and operates the data marketplace, facilitating the exchange of stable data products across users and tenants. We motivate and show how we run scalable decentralized data governance. 

All of this is built and run for Cimpress Technology (CT), which operates the Mass Customization Platform for Cimpress and its businesses. The CT data platform serves 1000s of users from different platform participants, with data sourced from heterogeneous sources. Data is ingested at a rate of well over 1000 individual messages per second and serves more than 100k analytical queries daily.
\end{abstract}


\section{Introduction}\label{section_intro}
The transition from multiple monolithic systems, that have traditionally relied on shared datastores for exchange of information, to a microservice-driven architecture, has resulted in the proliferation of data across the organization. Suddenly, hundreds of developers generate data, each in their unique choice of data structures, on various cloud infrastructure and storage systems, in different programming languages and paradigms. 

This geometric growth of data generating sources has increased the appetite for analytics, both within a specific domain as well as to outside consumers. By definition, each microservice has a bounded domain context where data are not accessible to outside actors. The objective of exposing data for cross-domain analysis to drive value becomes a challenge. Building a data-driven organization and successfully adopting artificial intelligence in software products and business decisions needs foundations in stable, findable, and accessible data products\footnote{https://mitsloan.mit.edu/ideas-made-to-matter/how-to-build-a-data-driven-company}\footnote{https://sloanreview.mit.edu/article/the-data-problem-stalling-ai/}.

Cimpress rapidly adopted a microservice strategy providing APIs and components that enable the entire order lifecycle in the Enterprise Mass Customization~\cite{pine1993mass} Platform (MCP) case. The flow includes manufacturing, pricing, shipping, document processing, marketing and e-commerce with platform participants being merchants, fulfillers, buyers, and sellers. With each step of the process generating data, applicable to each platform participant, the requirement exists to provide a self-service, API first, multi-tenant \textit{data platform} that addresses the complete data journey from ingestion \& storage to curation, consumption and advanced analytics.

The data platform provides tooling, provisioned infrastructure, and light-weight human and systems processes to accommodate the data journey. It is important to note that the team building and operating the data platform does not own any business data and does not provide centralized support for data loads or curation beyond education and thought partnership. A team of fewer than 40 engineers is able to research, develop, and operate the data platform for a highly decentralized enterprise with revenue of almost \$3B.

A decentralized model, particularly for a business of this size requires a data marketplace which supports the flow of data products, their exchange, recommendations, documentation and data governance. In the case of the \textit{gourmet data product marketplace}, it is mostly distributed, light-weight data governance that focuses on increasing the trust in the data products. At the same time, the data platform supports lean onboarding through a seamless data product publishing process that does not have bottlenecks in centralized data governance functions and allows owners and consumers of data to classify and distinguish raw, experimental, test or internal data sets from fully production-ready, fully supported data products.

\subsection {Business Context}
The primary culture shifts on providing and consuming data in a microservice environment helped shape the requirements for the data platform. Key challenges were:
\begin{itemize}
\item Cimpress is a group of autonomous businesses with full ownership and authority over their operations, including data analytics. With the shift to decentralized teams, each owning domain-bound application built microservices. This resulted in the fragmentation of data sources and unclear ownership.
\item The immense quantity of independent services and requirement of deep domain knowledge excluded the option for running a centralized data team. This shift impacted all data stakeholders; consumers no longer have access to a single, centralized data team responsible for data engineering, data curation, and data modeling.
\item Data producers, who in the past only exposed their datastores, became responsible for providing data in a push mode in addition to treating data like they treated their APIs - with all the quality expectations and service levels that come with it. Teams are responsible for the entire journey of the data.
\end{itemize}

The main data platform actors are \textit{data producers} who produce and ingest the data and \textit{data consumers} who make use of the data in the data platform. These use cases drove the \textit{data platform} to become an opinionated infrastructure that provides a distributed marketplace where anyone can create stable datasets and build data products for a broad set of business needs.

A data product, according to our internally-used definition, is \textit{"An objective-based product, whose primary function is to use data to facilitate the end goal"}.
We consider data products to have high levels of service commitments, adoption metrics and well-defined roadmaps. On the technical layer, data products can contain one or more of the following data resources:

\begin{itemize}
\item Stable Datasets
\begin{itemize}
\item That conform to STABLE qualifications, with the intent to be shareable.
\item \eg Can be raw, derived or 3rd party data. As opposed to internal data.
\end{itemize}
\item Algorithms
\begin{itemize}
 \item \eg predictions, data masking, customer value calculation
\end{itemize}
\item Decision support systems
\begin{itemize}
 \item \eg A data warehouse or logical data model 
\end{itemize}
\item Transformation or curation logic
\begin{itemize}
 \item \eg A collection of data transformation code, optionally with modeling around a visualization layer
\end{itemize}
\item Visualization components
\begin{itemize}
 \item \eg A component paired with the relevant logical data modeling
\end{itemize}
\item Automated decision making
\begin{itemize}
 \item \eg recommendations, calculation of data sets as a service
\end{itemize}
\end{itemize}

\subsection {Why Data Platform and Marketplace?}
The data platform leverages economies of scale on a shared infrastructure aligning on a common technical stack, allowing for data consumption without data movement. It provides common security, authorization, and authentication frameworks. Considering that data producers, and consumers have diverse motivations, expectations, and requirements while working in the data ecosystem; a data marketplace facilitates these data personas to focus on delivering value-added activities, building data services and products.

For instance, \textit{Data producers} do not have to invest effort in managing and maintaining data pipelines to ingest the data, as they can use self-service, low barrier, scalable, easy-to-use, low-maintenance options to push data. Also, data producers need to transform and curate their data to make it useful, stable, and discoverable by others. Data producers need to allow others to consume their data and further build on top of it with quality and trust expectations. The data product exposed in a discovery and cataloging services is findable by the potential consumers. A platform evaluates the quality and stability of the data to set up the consumers' expectations. The definition and guidelines on creating stable data sets are aligned across actors and tenants. The platform has modules to manage access and data sharing across data lakes, regions, and tenants. Furthermore, data producers want to have code-first, SQL-based, or modeling-based data transformations and representations that can be versioned, easy-to-test, documented, and transparent. They look for the same qualities of self-service and low barriers in modeling data without prerequisite requirements on materializing data to fit dimensional and fact models that frequently require batch processing. 

\textit{Data consumers} need to find useful, trusted, and stable data products. A data marketplace evaluates the data sets and promotes data products' stability, and aggregates user feedback into recommendations. Data consumers need to share data further with their customers, whether they are part of their team or organization or work for other data stakeholders.

\vspace{5mm}

The remainder of this paper is organized into distinct sections. In Section~\ref{section_data_platform}, we describe the self-service multi-tenant data platform: its design principals, architecture, implementation details, technology, and provider choice. Section~\ref{section_data_platform} is divided into subsections corresponding to platform modules. The order follows the data flow and moves from the basic services that made up the platform to the ones that build its marketplace properties, enrich automation and usability. In Section~\ref{section_marketplace}, we describe the gourmet data product marketplace, and we show why the data product marketplace needs to be built as a part of a data platform. We show it on the Cimpress use case. In Section~\ref{section_related} we go through the related work on data platforms and marketplaces. Finally, we conclude the paper in Section~\ref{section_conclusion}, which also reveals the development and research plans.

\section{The data platform}\label{section_data_platform}

\begin{figure*}[ht]
 \includegraphics[scale=0.19]{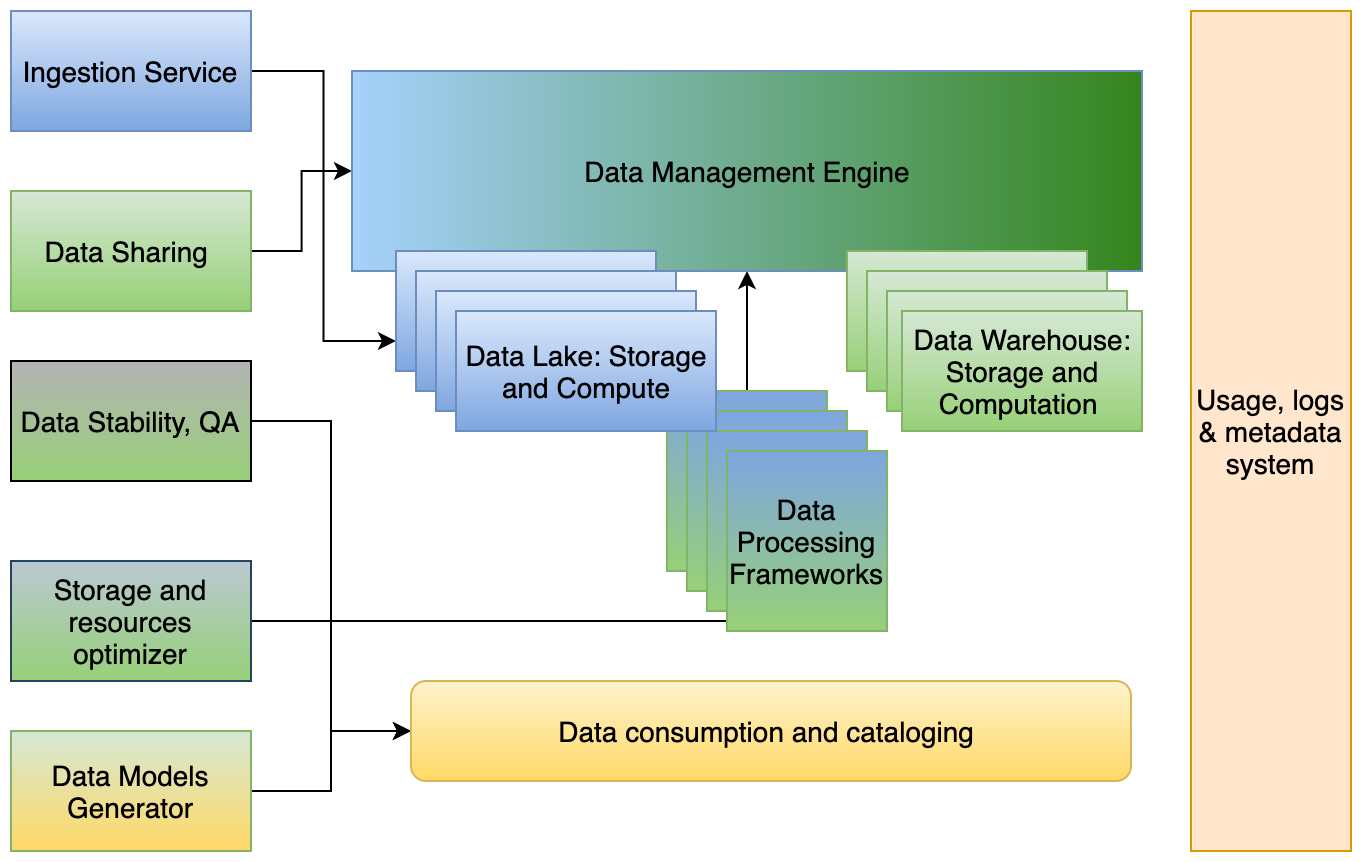}
 \centering
 \caption{A schema shows the main modules of the data platform in a multi-tenant setting. Each tenant has its dedicated Data Management Engine that comprises data lakes and data warehouses (DW). Data is ingested into a data lake. DW and data lakes have storage and compute resources and can be used independently. Both engineers and analysts control the data and data services through RESTful APIs. Data itself is accessible through the consumption layer or directly in the data lake or data warehouse environment. Arrows show the modules act on different data platform layers. The sensing system acquires all interaction with the services and data. All modules literally use this system.}
 \label{data_platform_diagram}
 \end{figure*}

The data platform provides cloud-based scalable self-service capabilities to ingest, store, curate, explore, and exchange data sets. These capabilities enable the organization's analytical data flow that ends with different data consumption models, \eg reporting, near-real-time dashboards, real-time recommendation models, data science, and machine learning. In Figure~\ref{data_platform_diagram} we show an overview of the data platform and its modules. The details of each module are described in the subsections. In Table~\ref{table_label_modules} we present the primary services that make up the platform.

\begin{table*}[!htbp]
\centering
\caption{Table with a list of the most important multi-tenant modules making up the gourmet marketplace}
\label{table_label_modules}
\begin{tabular}{| m{7cm} | m{10cm}|}
\hline
\textbf{Module - Service name} & \textbf{Role} \\ \hline

Ingestion Service - River & RESTful Data Stream Management and Direct Event Ingestion \\ \hline

Data Management Engine - PDW & Data Warehouse and Data Lake  \\ \hline

Data Sharing - DUCK & Data Exchange Management \\ \hline

Data Models Generator - Transformer & Document schema,\eg JSON to data models transformation  \\ \hline

Monitoring - SIFT & Log and metrics acquisition, processing, and reporting \\ \hline

Data Lineage & Data sets, users, and applications dependency discovery \\ \hline

Monitoring, resource optimizer - DMON & Data Sets and resources optimization recommendations \\ \hline

Access Management & Self-service authorization \& authentication \\ \hline

Data Marketplace - Data Discovery & Data set search engine and cataloging \\ \hline

Data Marketplace - Data Stability Evaluator & Evaluates \& Classifies data products based on their performance, quality, and metadata \\ \hline

Data Processing - Advanced Analytics and Modelling & Spark, DBT, Databricks \\ \hline
 
Data Processing - Orchestration & Matillion, Airflow, Snowflake Tasks \\ \hline

Compliance - PIIScanner & Automatic detection of PII in the entire DW \\ \hline

Reporting - Looker & Data analytics, reports, and visualization  \\ \hline

\end{tabular}
\end{table*}

\subsection{Design Principles}
The data platform builds on the following set of principles driving the design choice for technologies, decisions of build \vs buy for different modules, technology providers, and runtime cloud environment: (1) At the core, the platform acts as a service to developers, analysts, data engineers, and data scientists. Its deployment can be replicated for any other organization without resolving monolithic dependencies or specific business domain problems. (2) The platform addresses multi-tenancy in all the modules, without operational cost or usage constraints on the number of tenants and its users. Also, there are no overheads in cost or infrastructure in onboarding new tenants. (3) The data platform is technology agnostic.

Based on these principles, the data platform itself is API-first and self-service for its users. All modules have APIs exposed that enable creating code-controlled data products. Data owners, typically software engineers, extend their software practices with flexible-to-use APIs and tools that fit their CI/CD and other operational needs. Data scientists or data engineers are enabled for code-first, CI/CD-based development and deployment practices.

Being agnostic to technology makes it flexible for various integration scenarios. All aspects of our data platform can be controlled and configured via RESTful APIs. They enable integration with a diverse set of microservices or third-party services to ingest data. It avoids limiting the data flow to a single programming language, \eg often Python, or system, \eg limiting CI/CD integration to a single SaaS tool like GitLab.

The data platform components are a cloud-native combination of APIs, user interfaces, and managed services combined with various SaaS tools. All infrastructure is scalable to current and future needs, typically with both horizontal and vertical scaling. Monitoring and automation allow keeping costs in control and, in many cases, continuously drive down. From a compliance and observability perspective, all activity is logged, and most of this logging data is provided to users of the data platform for self-service diagnostics and root cause analysis.

\subsection{Data Management Engine}

Each tenant has its Private Data Warehouse (PDW). A tenant can be a separate business or a 3rd party collaborator on the data platform. A PDW consists of a Data Management Engine comprising a data lake and a data warehouse that is logically separated in the engine by having different resources and storage zones. Each PDW has its storage \& computing resources to serve the analytical workloads, \eg~data ingestion, transformation, aggregations, pivoting. Currently, the data platform uses Snowflake~\cite{snowflake_paper} as the Data Management Engine leveraging the premise of storage and compute resource separation. Snowflake was chosen as it meets the data platform design principles; it enables the creation of virtual compute warehouses with zero overhead in maintenance and cost. The warehouses automatically scale horizontally with the expected load and scale vertically during the runtime without query disruption. The primary interaction between users and a data warehouse happens through SQL, or its abstractions through third party connectors, \eg~Databricks - Spark data frames, AWS Glue. Through the use of a common storage engine, separated logically for data lake workloads and data warehouse workloads, direct query, and utilization of data sets for both use-cases is accommodated with zero data movement or complex transformation and loading processes.

Asynchronous data sharing is possible, with zero data movement, across all tenants on the platform through logical pointers in the data facilitating read-only access to shared data sets in a single environment. It allows the combination of both owned and shared data.

A custom, specialized module responsible for data sharing, named DUCK, proactively verifies the compatibility of shared data changes avoiding breaking changes and conflicts. Also, each tenant can create \textit{read-only PDWs} that can be used be used by trusted third parties, including a configured set of shared data.

\subsection{Simple RESTful Data Ingestion}
\begin{figure}[!htbp]
\centerline{\includegraphics[scale=0.05]{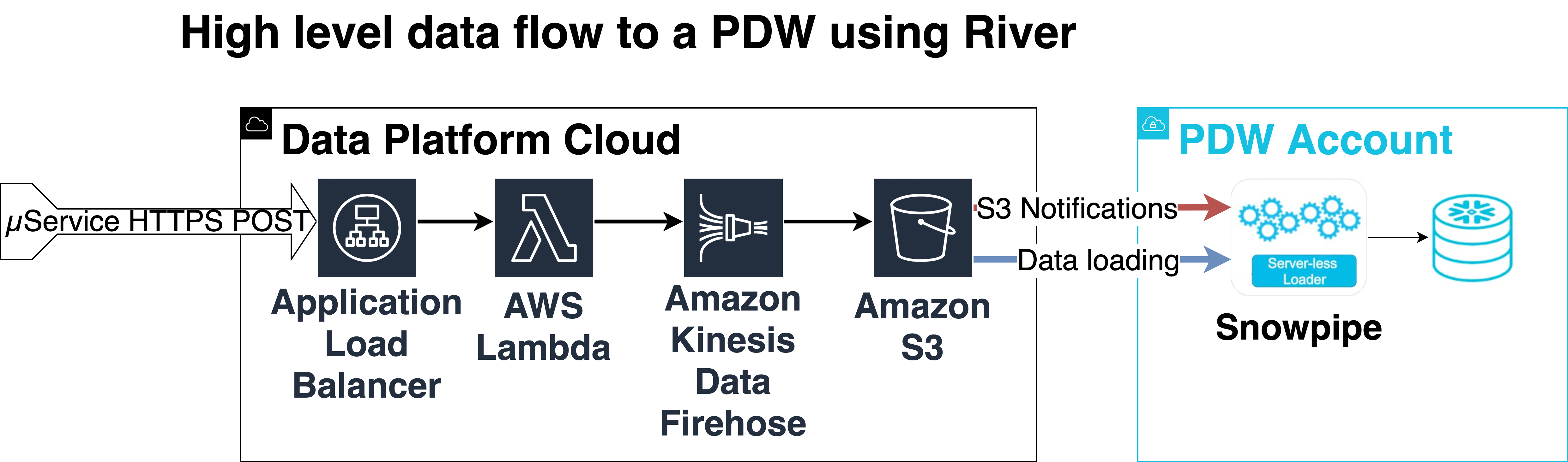}}
 \caption{Cloud and PDW components that make up River. Arrows indicate the data flow.}
 \label{figure_river}
 \end{figure}

River is the data platform's self-service data ingestion service, see Figure~\ref{figure_river}. Data typically comes from software applications, where it's sent in parallel to a particular application's transactional storage systems. The service is built to be developer-first and API-first. A single end-to-end ingestion pipeline is a \textit{River stream} with a fixed name and a data lake destination - a table in the PDW, \ie base table that stores the data without any transformations applied. Streams are organized in namespaces that correspond to databases and schemas in the data warehouses. Setting up a stream can be achieved easily through an intuitive web user interface. Every River stream has its endpoint for data ingestion and its serverless infrastructure. One can instantly send data to a PDW by posting data in a valid JSON payload to the created stream endpoint. Data is routed to the appropriate tenant in our multi-tenant architecture, and access to the data can be set up at stream creation. River provides capabilities of data versioning which is a feature that filters out only the latest version of a field from the raw data, and deduplication which removes duplicates from the raw data, leveraging CDC (Capture Data Change). River also applies clustering on the river base/raw tables to optimize query performance, \ie maximize partitions prunning. 

\begin{figure}[!htbp]
\centerline{\includegraphics[scale=0.34]{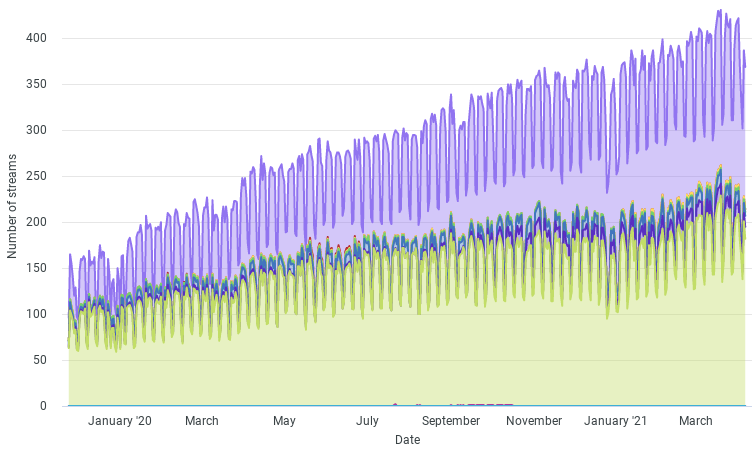}}
 \caption{Graph showing the number of daily active streams for all PDWs for the last 1.5 years. Purple and green stand for the two largest tenants. The number of active streams grew almost 4x times in this period. }
 \label{figure_river_api_calls}
 \end{figure}

As data is sent in a JSON format through the microservices, it is stored in a Variant type column in a PDW. Pushing the data to the platform is achieved by calling a River stream endpoint with a POST request. The data pushed to the microservice is first picked up by AWS Kinesis Firehose stream that loads this data to the appropriate tenant staging zone, \ie S3 bucket with a date-partitioned directory for each River stream. There is an external stage in the data lake in PDW that references the loaded stream files in the staging zone, \ie everything below the tenant-stream-id directory. We leverage serverless data load, \ie Snowflake Snowpipe, that ingests the S3 data physically from the external stage into a PDW table as soon as it appears in the stage. The load is asynchronous and based on event notifications from cloud storage on new data files' arrival. Snowpipe copies the files into a queue, from which they are loaded into the target stream table in a continuous, serverless way. Currently, River processes about 1000 events per second on average on a typical weekday, with multiples of that at peak times. There are about 1000 streams created with about 500 daily active ones. See Figure~\ref{figure_river_api_calls} for the stream adoption and activity.

\subsection{Automatic Data Transformation of Data Models for Semi-structured Data}
We promote \textit{schema-on-read} and Extract Load Transform (ELT) paradigms - data is not transformed before the load to the warehouse. This paradigm enables to experiment and iterate fast on the optimal transformation logic from semi-structured to tabular. Consequently, data lake and data warehouse store data sets with various structures, \eg tabular, semi-structured, JSONs, geospatial. To transform any semi-structured data, a user needs to explore the keys and nested objects. It is a repeatable process that can be automated to complete 90\% of the job for a human. 

The data platform built Transformer. Transformer is a microservice that uses semi-structured data \eg~JSON as input and generates 1) data models, \eg~Looker LookML Views and Explores, and 2) SQL that enables to explore the nested data. The parsing engine is released as an open source library\footnote{https://github.com/Cimpress-MCP/j2v}. The algorithms automatically detect nested data structures (without limits on the nesting levels), certain data types, provide NULL handling policies, and resolve key-path name conflicts. The service enables to explore the semi-structured data in visualization tools and take the generated SQL to set up a transformation pipeline in seconds from the first data being ingested.

\subsection{Adoption, Monitoring, and Usage: Sensing System: Logs, Metrics, Queries, and Query Plans}

Data producers and data consumers should see everything significant to the data platform's data product usage. Also, exposing resource and usage cost as show/chargeback at granular levels, e.g. per each user is a primary capability of any multi-tenant platform - tenants need to know what they are billed for. PDW administrators and data engineers should measure and perform root cause analysis of any resource consumption. Data product providers are accountable for performance and stability improvements. Common problems of diverse data personas are (1) Who uses my data? Is it used correctly? How is it joined with other data? (2) How do I leverage the data platform to share the data with the PDWs? How do the businesses use the data? (3) What if I change the data structure, who and how may they be affected? (4) How is my data set performing? How can I improve it? (5) Which columns and JSON fields are frequently used?

The data platform empowers users to analyze, audit, log, and troubleshoot the whole data life cycle and used resources. Such analysis and discovery is made at various entity types like data products (including all types listed in the \textit{Introduction Section} ), compute and storage resources, and PDWs. Examples of the analysis and discovery are understanding the means of how a data product is consumed, auditing for access, identifying the usage patterns, auditing and validating the configuration, investigating and introducing data product structure changes.

We created a data product, SIFT, which stands for \textit{sift - examine (something) thoroughly so as to isolate that which is most important}. SIFT collects all usage data, \ie query history, query execution plans, all computational, storage, and network resources configuration and usage, data pipelines and jobs, access, sessions, cost, coming from all the tenants, processes and analyses it, and stores it in a centralized platform sink. This data product is exposed to users as APIs, visualizations, and directly by accessing the data sets on the data lake or in a specific PDW through the data exchange. SIFT data is also a data source for the data platform modules like performance \& cost optimizations and data lineage.

\subsection{Data Object Dependencies as Query Graph Unions}
\begin{figure}[!htbp]
 \centerline{\includegraphics[scale=0.26]{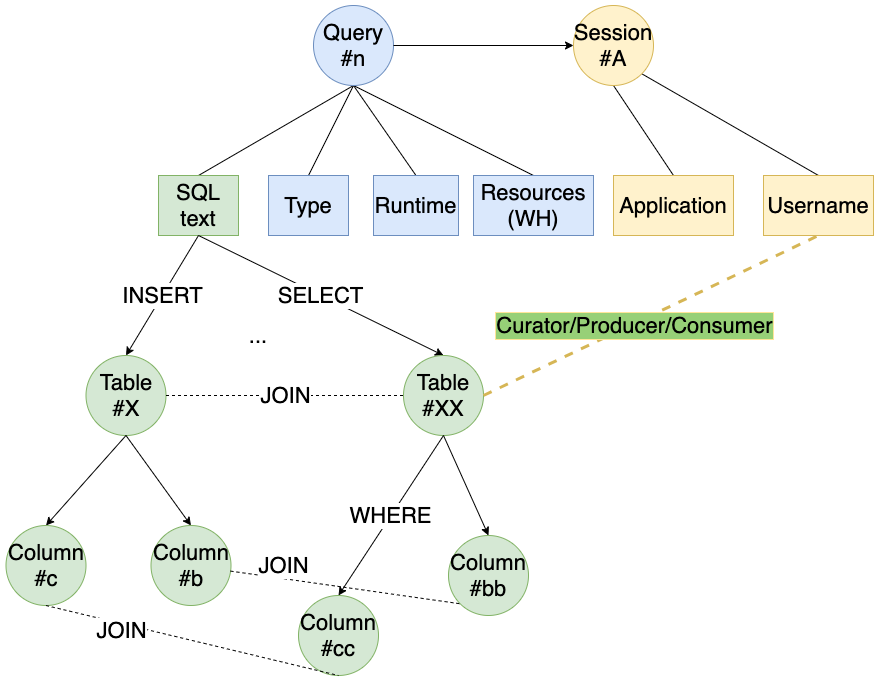}}
 \caption{An example graph presenting some of the information avaiable for each query being executed: breakdown of the data sets dependencies, session, runtime and resources data. Doing a union of all these graphs makes up the preceise dependencies between users, data sets, and resources with details on each.}
 \label{query_data_schema}
 \end{figure}

The usage data coming from SIFT is fundamental to resolving all the dynamic dependencies - revealed in the runtime and static ones - known from code dependencies between data objects, users, PDWs, and resources (see Figure~\ref{query_data_schema}). Resolving these dependencies enables rich data discovery service, data product, and resources change management, proactive monitoring of the breaking changes, and efficient root cause analysis. SIFT has a query text, query plan, and other information for each query ever run in the system. 

\begin{figure}[!htbp]
 \centerline{\includegraphics[scale=0.31]{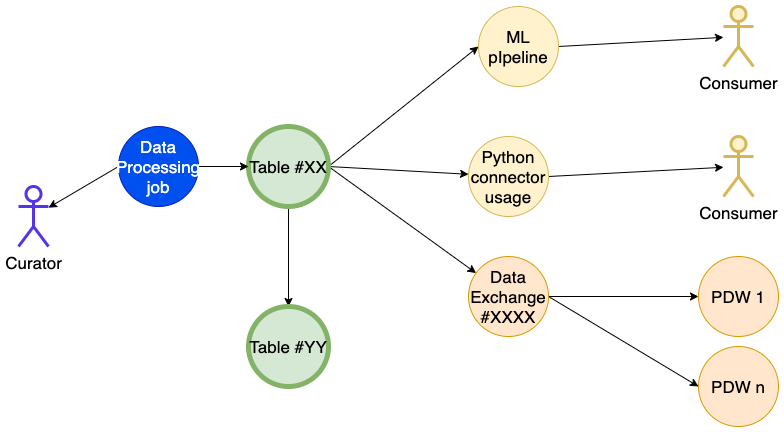}}
 \caption{An example graph showing the dependencies of some table, the graph is a result of many graphs unions, captured in a given time window.}
 \label{data_lineage_example}
 \end{figure}

We combine the data coming from the query plan with the data coming from parsing the query text to resolve the multi-level nested dependencies, such as views, data shares, or data from other locations, \eg scanning both data lake and data warehouse in the same query. Then, we combine all the graphs to create a data lineage data that shows dependencies in the entire platform (see Figure~\ref{data_lineage_example}). The result allows for social graph analysis of the components \eg looking for node clusters and cluster connecting nodes. As every information has a timestamp, any dependencies can be tracked and aggregated in a defined time window. It gives the capability to explore user interaction changes with data over time, adoption of various data analytics and machine learning applications, and overall complexity of the data product dependencies.

\subsection{Towards Self-healing System: Automatic Storage and Data Transformation Recommendations}

Whether offered as a platform or as a service, any data storage computing system has to deal with trade-offs to achieve maximal performance to increase the user satisfaction and reduce cost of cloud resources. Although tenants leverage the usage data to optimize for the above factors, relying only on them to optimize their resources for cost and performance is not scalable and sustainable at the platform level. Also, it works in contradiction to the economics of scale that such a platform should leverage. Optimizations and recommendations impacting the resource consumption happen at many levels of the technology stack, \eg cloud provider insights, computational resources scaling policies, SQL compiler optimizations. 

The data platform has a service, named \textit{DMON}, that analyzes the usage data and triggers optimization, keeping data producers in the feedback control loop. Advising the data owners on the optimizations is a scalable solution with an additional function of providing the learning, and the empowered data producer executes the final action. For instance, a data owner automatically receives a notification with the precise clustering suggestions for a data set. A user accepts or rejects such suggestions, providing input to the semi-automated system. Another example is the detection of overusing of the data management system's failsafe and recovery policy at the data set level. DMON detects data sets derived from others through data processing jobs, and in most cases, such data sets can be recalculated at any time, so there is no need to store backups of them and their snapshots. 

The long-term objective is to tune the advising system and apply these suggestions to data owners automatically. To do that, we need to lower the rate of false positives and errors in the suggestions themselves. The way to achieve it is to clone the data sets into the prepared testbed system and run a workload that mimics the usage pattern.

\subsection{Data Processing and Orchestration}
The data platform provides core capabilities of data transformation, engineering, and orchestration of data processes. We leverage DBT\footnote{https://github.com/fishtown-analytics/dbt} that uses the code-first approach to maintain transformation logic using version control. DBT also allows for visualizing the dependencies between the data models and quality assurance, \eg unit testing on data transformation. For instance, we empower data producers to create stable data sets through code-controlled templated projects based on DBT and sample API calls to the data platform services. All these enable data producers to be set up to create their data sets - data products within seconds and align the standard between the tenants. Apart from the platform's capabilities, each tenant is free to use their processing and analytics toolings, such as Spark, Databricks, and stream analytics.

The data platform enables to orchestrate data processes. A platform capability is built-in data management scheduling (Snowflake Tasks) and Apache Airflow. There are also external tools such as Matillion\footnote{https://www.matillion.com/} that combine orchestration and transformation and have a low entry barrier to data engineering.

\subsection{Data Privacy - Automatic PII Detection}
Users of the data platform store data sets that contain personally identifiable information (PII). The majority of such data comes from eCommerce, marketing, and customer personalization services, algorithms, and machine learning models. That is why we built automatic PII detection and PII metadata management. The module runs automated scanners across the PDWs to detect and flag PII fields. This information is extracted from metadata and deep data sampling by a version-controlled set of rules regex and spaCy\footnote{http://spacy.io/} NLP models for entity recognition. In the end, a user receives notifications and can verify the results of the periodical scan.

\subsection{Access Management}
The data platform provides self-service API \& UI to manage access to the data products. Users can leverage the tool to grant access to the data products they own or co-manage. Self-service tools put the control into the hands of data owners and enable decentralized and scalable access management. All PDWs have a small number of administrators to orchestrate and grant advanced privileges of any data product and data object. The tool allows to create PDW users (both human and machine2machine) and roles, and allows to grant various privileges to the managed data products. The access management is integrated with the enterprise self-service centralized authorization system of the organization. Each tenant can use an access group that has defined members, admins, resources, and permissions on these resources for the access management. The access groups can also be configured to manage cross tenant accesses. For instance, it is possible to grant access to non-interactive users in various PDW data warehouses, using only a single access group.

\subsection{Data Quality and Stability Services}
Enterprises producing and consuming data in various business domains need data quality and governance management strategy and deployments~\cite{Wende2007AMF}. One of the biggest challenges in a decentralized data platform is data governance. In a self-service data platform, there is no human judgment through a central BI or analysts team, whether a particular data product meets the requirements to be released and considered as \textit{stable} or \textit{golden}. The responsible humans for data products are its producers, and each tenant can establish its own governance strategies. On one hand, having a centralized data governance function, \eg a council, ensures alignment of data sets across all data and its metadata. This keeps the data's stability and quality at the same level across the tenants. On the other hand, it slows the adoption of a platform and puts the importance to the process and formalisms over the value of a created data product. Opposite to it, a data governance structure that scales, keeps the ownership of data sets in the data domains, and enables the creation of modular data products. This decentralized approach is also known as a \textit{data mesh}\footnote{https://martinfowler.com/articles/data-mesh-principles.html}. 

The data platform needs a solution that guards data sets and data products' stability to scale the data sets into a high-quality data mesh. These are the main reasons we decided to provide built-in light-weight data governance through automatic evaluation of data products. The stability in the data platform covers the following metadata and data product categories: (1) \textit{Ownership \& Support} - the owner takes responsibility for the data and fixes it or works with the upstream providers. Consumers know how to ask for support. (2) \textit{Naming, Description, Bussiness Objective} - The documentation explains what for and how the data product is created. It reflects its meaning, data value delivered, and encourages exploration. (3) \textit{Read optimized access} - data products are clustered by tenant accounts when shared to tenants and can be quickly previewed, \eg takes seconds to sample the data set. (4) \textit{Addressability} - each data entity has unique address across the platform. Each of the above four categories has specific attributes evaluated on a data product by an automated service. Depending on how many of these attributes the data set meets, it is classified into one of the following categories: (A) Stable, (B) Investigable, or (C) Internal.

\section{Gourmet Data Product Marketplace in Mass Customization Platform}\label{section_marketplace}

The data platform facilitates a thriving data marketplace in which data producers and consumers transact. These transactions happen inside a platform tenant or across multiple tenants. The data marketplace enables the following workflow: get data in, curate it, make it a stable data product and transact. The marketplace is self-organizing, meaning that it promotes stable data products by usage and consumers' feedback. The marketplace is a hub for the exchange and consumption of enterprise data products, it links various data consumption models, protocols, clients, and applications.

Currently, the platform is used actively by 10 tenants, with 2 of them responsible for almost 70\% of usage and the rest distributed evenly. All tenants store about 200TBs of data in about 25k tables in their PDWs, including data sets of all types: staging, internal, unstable, stable. Data through River is ingested at a rate of well over 1000 individual messages per second. The platform serves 50k-100k analytical queries daily. 

Tenants create data products of various types (listed in Section~\ref{section_intro}), such as stable data sets, Machine Learning models, visualizations, shared algorithms, and reports. There are about 200 stable data sets transacted between tenants on the marketplace.

\section{Related Work}\label{section_related}
This section is dedicated to exploring the related work in data analytics platforms and data marketplace areas. In the gourmet marketplace, we intend to couple these areas and show how a marketplace can be "the mature" data platform stage. Consequently, we skip the review of the state of the art of particular technology choices and solutions for the platform components. We find it challenging to discover related work that reveals the data marketplace that considers the entire data product lifetime starting with the data production and ending with the data transaction in a marketplace.

\subsection{Data Marketplace}
The data marketplace solution is known since data exchange between parties delivered business gains, and the roots are 100 years ago at the stock exchange markets. In the 21st century, a data marketplace means a data provider publishes the data set, and a data consumer uses~ it\cite{stahl2016classification}. Recently data marketplaces are rapidly growing, caused by the need for data monetization and open data movements. Technically, the cause lies in the growth of systems generating enormous data sets, like the Internet of Things, edge computing, smart cities, and social media~\cite{smart_cities_marketplace,open_data_marketplace}. The growth of Machine Learning applications also made the need for real-time exchange of the commercial and open curated data set used for model training~\cite{ml_data_market}. 

The community highlights the meaning of organizing data teams, data governance and data stewardship, and ownership functions to create data products delivering actual business value~\cite{lit_review_governance, LARSON2016700}. A successful decentralized and multi-tenant data marketplace platform for data sellers and buyers consists of robust (1) data cataloging: query and search, metadata information; (2) data curation that improves the data quality and usability in response to user demand and requirements; and (3) recommendations, ratings, feedback that runs the discovery process of stable and valuable data sets. Important to note, the authors of a conceptual data governance framework and study review conclude with the importance of breaking the data silos in the organizations and finding the optimal data governance that keeps the balance between firm performance and data governance level~\cite{data_governance_concept_2019}. Another study of data governance research and technologies points to a gap in having the solutions that actively monitor the data processes and data quality~\cite{data_governance_literature_overview}. In the gourmet data marketplace, we handle data product exchange that is not limited to a particular type. We use the marketplace as the centralized and automated data governance. Still, we leverage the leading utility of serving a marketplace that is connecting consumers with producers.

\subsection{Data Analytics Platform}

We can observe many industrial applications of the data platforms, manufacturing systems, geospatial analytics, power grids, and eCommerce\cite{spacedataplatform,geo-spatial-platform,smart-grids-platform, manufacturing_big_data}. Enterprise data platforms have layers for data ingestion, storage, a data lake, and data consumption, including further processing and analytics applications. A typical setup includes an event streaming platform, \eg Kafka; a distributed storage system, \eg HDFS, and various data processing, curation, and consumption modules, \eg Flink, Spark, HBase, Jupyter Notebook, Impala or their enterprise equivalents \cite{big_data_survey}. How the data is consumed depends on the business cases and may include machine learning models, predictive analytics, graph databases, etc. The data consumption's uneasiness and usability are directly linked with the business value the entire data platform brings to the organization \cite{clouds-scalable}. 
For instance, the authors of \cite{spacedataplatform} describe a complete cloud-based data platform architected with microservices. It can operate on various data sources and has many data processing capabilities, \eg batch processing, stream processing. At this background, the data platform proposed by us has modules that optimize the data usage, allow for multi-tenant data exchange, and shorten the path from the data ingestion to data exploration. Another example of a data platform is presented in \cite{realtimeedge}. The proposed system is oriented for real-time data analytics using edge computing resources. The authors implement the control of users on the platform resources through contracts that contain QoS.
One of the recent papers on building and operating a hybrid datalake system running both on the public and private cloud - enterprise data platform is presented by~\cite{BAUER2021100181}. The authors extensively describe their architecture choice, experience, evolution, and current state of the platform. Interestingly, they point to the need for automation of every workflow to scale the platform correctly. Some of the technology choices, \eg Kafka, Nifi are the ones that we went through and changed for cloud-native technologies, \eg AWS Firehose. To compare the two platforms' sizes using the presented platform's usage data, we have about 10x higher data throughput and the number of users.

\section{Conclusion and future work}\label{section_conclusion}

This paper presented the self-service enterprise data marketplace as a solution for data-driven organizations to build and exchange data products. We described the challenges and principles of adopting and designing a data platform in a distributed microservice-oriented software company. Then, we presented the main modules of the multi-tenant platform responsible for the entire data lifecycle. The modules are created to reduce the technical barrier to enter the platform by diverse users such as data producers, and consumers. One of the most critical aspects of the presented data marketplace includes automated data product evaluation, which realizes a humanless data governance strategy.

The marketplace successfully operates in the enterprise organization, and it has no central team of data engineers responsible for data jobs. All data product ownership is distributed across the teams who created them. The platform has 10 tenants that exchange more than 200 data products and are responsible for the ingestion of well over 1000 messages per second. The platform's annual growth is measured in tens of percent, with more and more tenants moving their legacy data products to marketplace ones.

Our plans for research and development of the data marketplace cover native support of machine learning model lifecycle and including them in data sharing concepts through sharing of endpoints. Also, we want to open a marketplace for sharing data streams. Another important development branch is the data toolkit, a platform capability for data engineering, data transformation, and materialization. Data toolkit will support the deployment of data contracts and robust versioning of schemas.

\section*{Acknowledgments}
We would like to distinguish many contributors who did research, architecture, design, development, and managed the product, apart from the authors of this article. All of them helped in shaping the data platform, influencing the spirit and culture of the gourmet marketplace setting it up for success. We are particularly thankful to Marc Jacobson, Philip Favaloro, Tianyou Luo, Chris Bova, Maryia Borukhava, Paschalis Dimitriou, Sergio Rocha, Santosh JP Kumar, Gaurav Katwal, Mohit Purohit.

\bibliographystyle{IEEEtran}
\bibliography{bibl}

\end{document}